\documentclass[manuscript]{aastex}

\slugcomment{Submitted for publication in ApJ letters}

\shorttitle{The progenitor of SN1999gi}
\shortauthors{Smartt et al.}

\begin{document}


\title{An upper mass limit for the progenitor of the 
Type\,II-P supernova SN1999gi 
\footnote{Based on observations made with the NASA/ESA Hubble Space Telescope,
obtained from the data archive of the Space Telescope Institute, which is
operated by the Association of the Universities for Research in Astronomy, 
Inc., under NASA contract NAS 5-26555.}}


\author{Stephen J. Smartt,  Gerard F. Gilmore, Neil Trentham, Christopher A. Tout, 
Colin M. Frayn}
\affil{Institute of Astronomy, University of Cambridge, Madingley Road,
       CB3 OHA, Cambridge, England}


\email{sjs@ast.cam.ac.uk}


\begin{abstract}

Masses and progenitor evolutionary states of Type\,II supernovae
remain almost unconstrained by direct observations. Only one 
robust observation of a progenitor (SN1987A) and 
one plausible observation (SN1993J) are available. Neither 
matched theoretical predictions and in this {\em Letter} we
report limits on a third progenitor (SN1999gi). 
The Hubble Space Telescope has imaged the site of the Type\,II-P
supernova SN1999gi with the WFPC2 in two filters (F606W and F300W)
prior to explosion. The
distance to the host galaxy (NGC3184) of 7.9\,Mpc means that the most
luminous, massive stars are resolved as single objects in the archive
images. The supernova occurred in a resolved, young OB association
2.3\,kpc from the centre of NGC3184
with an association age of about 4\,Myrs.  
Follow-up images of SN1999gi with WFPC2 taken 
14 months after discovery determine the precise position of the SN
on the pre-explosion frames.
An upper limit of the absolute magnitude of the progenitor is 
estimated ($M_{\rm v}\geq -5.1$). By comparison with stellar 
evolutionary tracks this can 
be interpreted as a stellar mass, and we determine an upper mass
limit of  9$^{+3}_{-2}$M$_{\odot}$. We discuss the possibility of 
determining the masses or mass limits for numerous nearby core-collapse 
supernovae using the HST archive enhanced by our current SNAP programme.

\end{abstract}



\section{Introduction}

%Position 10 18 16.66 +41 26 28.2 (IAUCIRC 7329)
%Discovery date 1999 12 09
%PC1 = 0.0455'' per pix
%WF  = 0.0996'' per pix 

Supernovae of Types\,II and Ib/Ic are thought to follow 
core collapse in massive stars at the end of their lifetimes. 
However the only definite and unambiguous 
detection of a SN progenitor is that of SN1987A in the LMC \citep{white87}, 
which was a blue supergiant \citep[B3Ia;][]{wal89}. The progenitor of 
SN1993J in M81 was possibly identified as a K0\,Ia star with some 
excess $UB$ band flux either from unresolved OB association contamination or 
a hot companion \citep{alder94}. Neither progenitor is consistent 
with the canonical stellar evolution picture, where core-collapse occurs 
while the massive star is an M-supergiant. There is an understandable 
lack of observational data to constrain the last moments of stellar
evolution and explosion models for core collapse SNe. 
The supernova SN1999gi was discovered in NGC3184 on December 9th 1999
by \cite{nak99} at an unfiltered CCD magnitude of 14.5, 
and confirmed to be a Type\,II-P
\citep[see][and references therein]{leo2001}. Two 
SNAP programs with HST had previously imaged the 
SAB(rs)cd galaxy NGC3184 in two filters 
in June 1994 and 
June 1999 with the WFPC2. The distance to NGC3184 has been 
estimated by three methods which give good agreement of $7.9\pm2$\,Mpc 
($(m-M)_o=29.5\pm0.5$). 
The first is determined from the recession velocity of 
592\,kms$^{-1}$ (as given in NED and assuming 
H$_{0}=75$\,kms$^{-1}$Mpc$^{-1}$), 
the second from the tertiary calibration of \citet{devauc79}, 
and the third from the Type\,Ia calibration of \cite{pierce94}.
The relative proximity of SN1999gi, and the existence of high
spatial resolution observations prior to explosion alerted us to the 
possibility of identifying the massive progenitor star which exploded. 
This {\em Letter} describes the analysis of the pre-explosion 
data, showing that a progenitor star is not detected on either of the 
WFPC2 pre-explosion frames. We set direct observational limits on the 
bolometric 
magnitude and luminosity of the progenitor star (which we refer to 
as PSN1999gi) and by comparison with
stellar evolutionary models we can 
estimate an upper mass limit to the progenitor.

\section{Observational data and analysis}

Observations of the galaxy
NGC3184 were taken with the Wide-Field-Planetary-Camera-2 (WFPC2) on 
board the Hubble Space Telescope on two separate occasions
through the filters F606W and F300W
centred on 5957\AA~ and 2911\AA~ respectively (details
are given in Table\,1). The accuracy to which one can determine
absolute astrometry of the WFPC2 frames 
($\sim1-2''$) combined with the probable 
error on the position of SN1999gi ($\sim0.5-1''$), means that the 
supernova position cannot be accurately determined on the
pre-explosion data without a further calibration step. 
We hence re-observed 
the stellar association containing SN1999gi, with HST 
to locate precisely the supernova position on the 
pre-explosion observations. We placed the SN on WF4 to match
the existing data as closely as possible and took repeat
F606W/F300W data and supplemented this with F439W,
F814W and F555W observations (for future detailed analysis 
of the host stellar association). The supernova was also observed 
( in F555W only) by the SNAP programme 8602 to image the 
sites of recent nearby supernovae. In this case, SN1999gi 
was centred on the PC1 chip and, given the better effective 
resolution of this data, we have also used this image in our
analysis (see Fig.\,1). The SN originates in a resolved populous stellar
cluster which,  given the bright UV-magnitudes of the members, is 
clearly a large, young OB association which we will label
NGC3184-OB1. To match the pre-explosion
data to the pixel grid of the post-explosion 
frames (the F555W U6A05201R/2R is taken as reference), 
a simple 2-dimensional pixel transformation was calculated. 
Sixteen isolated, fairly bright stars common to all frames 
were identified and their centroids and magnitudes measured within the 
{\sc iraf daophot} package using PSF fitting techniques (see below for more
details). A geometric solution was calculated by fitting orthogonal
polynomials 
which resulted in residuals of $0.02''$ in each direction. Similar 
mean shifts between the stellar centroids in the resulting 
transformed frame and the reference F555W frame were measured
indicating excellent mapping accuracy, and we set the 
SN position at coordinates (0,0). 
At this position there is no detection of 
a single star either through visible inspection, finding routines 
or aperture photometry. There are two bright blue stars nearby
(OB1-1 and OB1-2) each $0.26''$ from the SN but both
are clearly visible on the post-explosion frames.

Therefore the progenitor is below the detection limits of both the
F300W and F606W pre-explosion data. Simulations with synthetic stars
were performed to determine the detection limits of the data in the
vicinity of the SN position. Model point spread functions (PSFs) 
were used to make two 3$\times$3 grids of stars of 
incrementally varying magnitudes
\citep[using Tiny Tim,][]{kris99} and these
were coadded to the observational
data frames. Aperture photometry was used within {\sc DAOPHOT} 
to find the stars and determine their magnitudes. We have taken 
the detection limit to be when more than 90\% of the synthetic 
stars are detected by the {\sc daofind} algorithm
at more than 3$\sigma$ of the background
level, their mean magnitudes are consistent
with those of the model PSFs 
and they are visually identifiable. 
We estimate the detection limit of the F606W exposures to be
$V_{\rm 606}=25.0\pm0.2$, and for the F300W exposures 
$U_{\rm 300}=23.5\pm0.2$. The calibration of 
\cite{whit99} was used to estimate the charge-transfer-efficiency 
effect, which reduces these limits by 4\% and 26\% respectively. 
Three other adjustments were made 
for geometric correction, aperture correction and contamination 
\citep{holtz1,holtz2}, giving final corrected limits of 
$V_{\rm 606}\leq 24.9 \pm0.2$, and $U_{\rm 300}\leq23.1\pm0.2$. 
Both these values
are consistent with the faintest detected objects
on the WF4 chip.

To estimate the reddening, and approximate age of the OB1 association,  
photometry of the single resolved stars was performed on the F300W and
F555W post-explosion exposures. Standard methods 
of point spread function (PSF) fitting photometry were used within
the IRAF package {\sc daophot} with TinyTim PSFS \citep[e.g.][]{john2001}.  
Counts were first determined within $0.2''$ apertures and the standard
corrections for geometric distortion, 
and charge-transfer-efficiency (CTE) were performed, according to  
\cite{holtz1} and \cite{whit99}. 
An aperture correction was then made to a $0.5''$ aperture, and 
contamination corrections in the two filters 
\citep[again][]{holtz1,holtz2} were applied. 
We removed objects which are much broader than a typical PSF and which have
poor goodness-of-fit parameters (using the sharpness and chi values 
produced by {\sc allstar}), and a  colour-magnitude diagram
is shown in Fig.\,2. 
The Geneva evolutionary isochrones from 
\citet{lej2001} were reddened assuming values of 
E(B-V)=$0-0.5$ by using transformations calculated within {\sc 
synphot}. A good fit to the luminous stars in the cluster is found with a 
value of E(B-V)=0.15, and an age of 4\,Myrs. 
Such a young age is required to fit the most luminous, and hottest stars in 
the cluster.  A value of E(B-V)=0.3
cannot be ruled out, although it does not trace the position of the 
$U_{300}$ brightest stars in the diagram. As discussed below, 
\cite{zar94} determined the reddening towards the H\,{\sc ii} region 
surrounding OB1 to estimate nebular abundances, and quote
A$_v$=1.07, more compatible with the E(B-V)=0.3 value. The best 
estimate from the stellar content appears to be E(B-V)=0.15, and we 
discuss the consequences for a larger reddening below. 

The detection limits of the $U_{\rm 300}$ and $V_{\rm 606}$ pre-explosion 
images can be converted to limits on the bolometric magnitude
and hence luminosity 
of the progenitor object, if we assume that the reddening estimated for the 
cluster stars is representative of that towards PSN1999gi. 
The $V_{\rm 606}$ upper limit can be converted to
an absolute visual magnitude through the standard relation (where 
$c_{\rm V-606} = m_{\rm V} - V_{\rm 606}$; a colour 
correction to convert the STMAG $V_{\rm 606}$ to a Johnson V, which is
dependent on stellar spectral type). 

\begin{equation}
M_{\rm bol} = 5 - 5 \log d - A_V + V_{\rm 606} + c_{\rm V-606} + BC \\
\end{equation}

As we have no knowledge of the spectral type of the progenitor we can 
calculate  $M_{\rm bol}$ for the range of supergiant spectral 
type possibilities, and
these are listed in Table\,2. The value for $c_{\rm V-606}$ is calculated 
for each spectral type within {\sc synphot} using the Bruzual atlas of 
model spectra provided, 
and the bolometric corrections (BC)
are taken from \cite{dril2000}. In this table we also 
list the upper limit of the stellar luminosity, assuming the solar 
$M_{\rm bol} = +4.74$. Similarly we can use the $U_{\rm 300}$ limiting
magnitude, and  through {\sc synphot} calculate the colour 
difference $c_{\rm V-300}$ as a function of spectral type, and hence 
calculate an equivalent $M_{\rm bol}$ and upper limit to the stellar
luminosity.

%% In this section, we use  the \subsection command to set off
%% a subsection.  \footnote is used to insert a footnote to the text.


%% This section also includes several of the displayed math environments
%% mentioned in the Author Guide.

\section{Discussion}
The limits we have set on the total luminosity of PSN1999gi
allow comparison with stellar evolutionary model predictions 
for pre-supernova massive stars, and 
we have chosen the Z=0.04 metallicity Geneva 
tracks \citep[with twice the normal mass-loss rates]{mey94,sch92}
for the following reason. 
The oxygen abundance gradient in NGC3184 has been determined by 
\cite{zar94} 
and their H\,{\sc ii} region $68''$N and $0''$E (2.3\,kpc radial distance)
from the nucleus of NGC3184
is almost certainly the H\,{\sc ii} region associated with NGC3184-OB1. 
The O/H abundance derived is 9.26$\pm$0.2\,dex,
suggesting that  the stars in this cluster are significantly 
more metal rich than solar \citep[8.83\,dex from][]{grev98}. 
The mass-loss rates used in these models are particularly high for the
more massive stars which experience Wolf-Rayet phases of evolution.
However for masses of less than
$10\,M_\odot$ (which we show below is the region of interest)
mass-loss plays only a minor role in
determining the point in the H-R diagram at which the star explodes
and we find similar constraints with alternative models, including those
with no mass loss \citep[e.g.][]{pols98}.
The WFPC2 pre-explosion images
are sensitive to all objects located in the shaded regions in Fig.\,3. 
For reference we show the position of Sk$-$69202, the B3Ia progenitor of 
SN1987A. Clearly PSN1999gi was not a similar massive
B-type progenitor, or it would have been detected on both the F606W
and F300W frames. 
The best, and fairly conservative, estimate of the upper mass of the 
progenitor is 9$^{+3}_{-2}$M$_{\odot}$. We have assumed a reddening 
of E(B-V)=0.15, and if this was increased to 0.3 (as discussed in the 
previous section), the luminosity limits would increase by 0.2. This 
effectively increases the mass limit to 12M$_{\odot}$, but still 
rules out a 15M$_{\odot}$ or above progenitor. 

This quite low initial mass value is surprising given the nature of the 
massive luminous stars in the OB1 association, and the age 
we derived from the 
\cite{lej2001} isochrones of 4\,Myrs.
The bluest, bright stars in this cluster 
have $U_{\rm 300}$  and $V_{\rm 555}$ magnitudes consistent with being 
very massive OB-supergiants  (40-60M$_{\odot}$) 
which have main-sequence lifetimes of $\sim4\pm1$\,Myrs. 
However a 9$^{+3}_{-2}$M$_{\odot}$
progenitor would have a lifetime of  $29^{-11}_{+19}$Myrs  
\citep{sch92}. The progenitor was clearly 
not one of the most massive stars in this association, in fact not 
even close in mass to the most luminous stars visible. This suggests
that the cluster did not form coevally and the age spread is 
 $\sim 25\pm 10$Myrs. Although quite 
broad,  this is not unprecedented; for example see the work on the 
the Milky Way double cluster h and $\chi$ Persei \citep{wild64,vogt71},
and the 30\,Doradus complex in the LMC \citep{wal97}. 
One caveat is that the detection limits 
are not sensitive to all evolved, hot massive stars i.e. Wolf-Rayet types. 
The WR stars span a large range in $M_{\rm v}$ 
\citep[$-3$ to $-7$;][]{vacca90}, and the fainter ones
would be below the detection limits, but given that SN1999gi was a
Type\,II-P containing substantial H\,{\sc i} it is unlikely that
the progenitor was a hydrogen depleted WR star. It could possibly have 
exploded during the early
stages of Wolf-Rayet evolution while it still has a significant
hydrogen envelope but is sufficiently blue to remain undetected.  However we
do not favour this possibility because it requires fine tuning of the
mass-loss rate but note that a similar evolutionary scenario would
explain the blue progenitor of SN1987a. One could 
alternatively imagine the progenitor being a high mass, dust embedded
star whose luminosity is extinguished by extinction, however
this dust would also have dimmed the SN event itself.
The lightcurve of SN1999gi peaked at approximately $V = 14^{m}$ 
(from various web sources), suggesting M$_v = -16$, which is not 
an underluminous event by any means \citep[see][]{miller90}. 

Our conclusion is that SN1999gi was not a
very high mass star and very likely had
an initial mass of less than 9$^{+3}_{-2}$M$_{\odot}$. The well maintained, 
and easily accessible HST archive has made this project feasible
and, in the future, will allow the investigation of SNe sites 
{\em before} explosion to be investigated in a systematic way. 
We have a Cycle\,10 SNAP project that will bring the number of 
late-type galaxies (within a distance of $\sim$17\,Mpc)
with WFPC2 2 or 3-colour photometry to $\sim$350. This 
will allow the sites of future core-collapse SNe in these
galaxies to be imaged prior to explosion and will 
extend the present work to a more statistically meaningful 
sample. The advent of Virtual Observatory initiatives both 
in Europe and the US (e.g. {\sc astrovirtel})
will make this type of project even 
easier to manage, and allow very fast reaction to events
where either limits can be set on the progenitor or 
in the event that a candidate star is identified.


%% Included in this acknowledgments section are examples of the
%% AASTeX hypertext markup commands. Use \url without the optional [HREF]
%% argument when you want to print the url directly in the text. Otherwise,
%% use either \url or \anchor, with the HREF as the first argument and the
%% text to be printed in the second.

\acknowledgments
SJS, NT and CMF acknowledge support from PPARC, 
CAT thanks Churchill College for
a fellowship. We thank the {\sc astrovirtel} initiative at ESO/ST-ECF
for new software development for searching HST/ESO archives.
We acknowledge A. Filippenko and the
team of HST SNAP 8602 for waiving the proprietary period
of their data. 




\clearpage

%% Tables should be submitted one per page, so put a \clearpage before

%% If you use the table environment, please indicate horizontal rules using
%% Do not put multiple tabular environments within a single table.
%% The optional \label should appear inside the \caption command.

\begin{table}
\begin{center}
\caption{Details of the HST WFPC2 Observational data. The location 
column indicates were the SN1999gi position was placed on WFPC2, 
with PC1 having a plate scale of $0.0455''$ per pixel, 
WF4 having $0.0996''$ per pixel. }
\begin{tabular}{lllcc}
\tableline\tableline
Dataset &  Date & Filter & Exposure Time  & Location  \\ \tableline
U3HY2901R/2R & 9th June 1994         &  F606W  & 160s & WF4\\
U29R1501T/2T & 21st June 1999        &  F300W & 800s  & WF4   \\
\\
U6BR0105R/6R/7R & 2nd March 2001     &  F300W & 1200s  & WF4    \\
U6A05201R/2R & 24th January 2001     & F555W &  700s   & PC1  \\
\tableline						  
\end{tabular}	
\end{center}
\end{table}
\clearpage

\begin{table}
\begin{center}
\caption{Limits on the bolometric magnitudes and luminosity of the 
progenitor from the $V_{\rm 606}$ and $U_{\rm 300}$ detection limits
on the pre-explosion frames.}
\begin{tabular}{lllcccccc}
\tableline\tableline
          &          &         &  \multicolumn{3}{c}{Limits from $V_{\rm 606}$}  &  \multicolumn{3}{c}{Limits from $U_{\rm 300}$}\\
 Sp. Type & $T_{\rm eff}$ (K) & & $c_{\rm V-606}+$BC & $M_{\rm bol}$ & $\log L/L_{\odot}$ &  $c_{\rm V-300}+$BC & $M_{\rm bol}$ & $\log L/L_{\odot}$ \\\tableline
  B0   &  28500 & ~~ &  $-$2.90 &  $-$7.96  &  5.08 &  $-$1.06 &  $-$7.92  & 5.06\\
  B3   &  18000 & ~~ &  $-$1.57 &  $-$6.63  &  4.55 &  $-$0.15 &  $-$7.01  & 4.70\\
  B8   &  13000 & ~~ &  $-$0.87 &  $-$5.93  &  4.27 &  $-$0.34 &  $-$6.52  & 4.50\\
  A0   &  11000 & ~~ &  $-$0.62 &  $-$5.68  &  4.17 &  $-$0.93 &  $-$7.79  & 5.01\\
  F0   &  7500  & ~~ &  $-$0.18 &  $-$5.24  &  3.99 &  $-$1.63 &  $-$8.49  & 5.29 \\
  G0   &  5370  & ~~ &  $-$0.19 &  $-$5.25  &  4.00 &  $-$2.28 &  $-$9.04  & 5.55\\
  K0   &  4550  & ~~ &  $-$0.47 &  $-$5.53  &  4.11 &  $-$3.51 &  $-$10.37 & 6.04\\
  M0   &  3620  & ~~ &  $-$0.99 &  $-$6.05  &  4.32 &  $-$5.33 &  $-$12.19 & 6.77\\
  M2   &  3370  & ~~ &  $-$1.27 &  $-$6.33  &  4.43 &  $-$5.62 &  $-$12.48 & 6.89\\
  M5   &  2880  & ~~ &  $-$3.11 &  $-$8.17  &  5.16 &  $-$7.47 &  $-$14.33 & 7.63\\
						  	       
\tableline						  
\end{tabular}					  
\end{center}						  
%						  
% Breakdown of Color corrections + BC		  
%						  
%SpTy c_{\rm V-606}    BC                   c_{\rm V-300}     BC   
%---------------------------------------------------------------------------
%
%
%O7                                          1.52           - 3.5  = -1.98 
%09.5                                        1.70           - 3.2  = -1.50
%B0   -0.27         - 2.63 =  -2.90          1.57           - 2.63 = -1.06
%B3   -0.26         - 1.31 =  -1.57          1.16     	    - 1.31 = -0.15
%B5                                          0.75           - 0.95 = -0.20
%B8   -0.21         - 0.66 =  -0.87          0.32	    - 0.66 = -0.34
%A0   -0.21         - 0.41 =  -0.62         -0.52	    - 0.41 = -0.93
%F0   -0.17         - 0.01 =  -0.18         -1.62	    - 0.01 = -1.63
%G0   -0.04         - 0.15 =  -0.19         -2.13	    - 0.15 = -2.28
%K0   +0.03         - 0.50 =  -0.47         -3.01	    - 0.50 = -3.51
%M0   +0.30         - 1.29 =  -0.99         -4.00	    - 1.29 = -5.33
%M2   +0.35         - 1.62 =  -1.27         -4.00	    - 1.62 = -5.62
%M5   +0.36         - 3.47 =  -3.11         -4.00	    - 3.47 = -7.47
\end{table}	

\clearpage

\begin{figure}
\epsscale{0.7}
\plotone{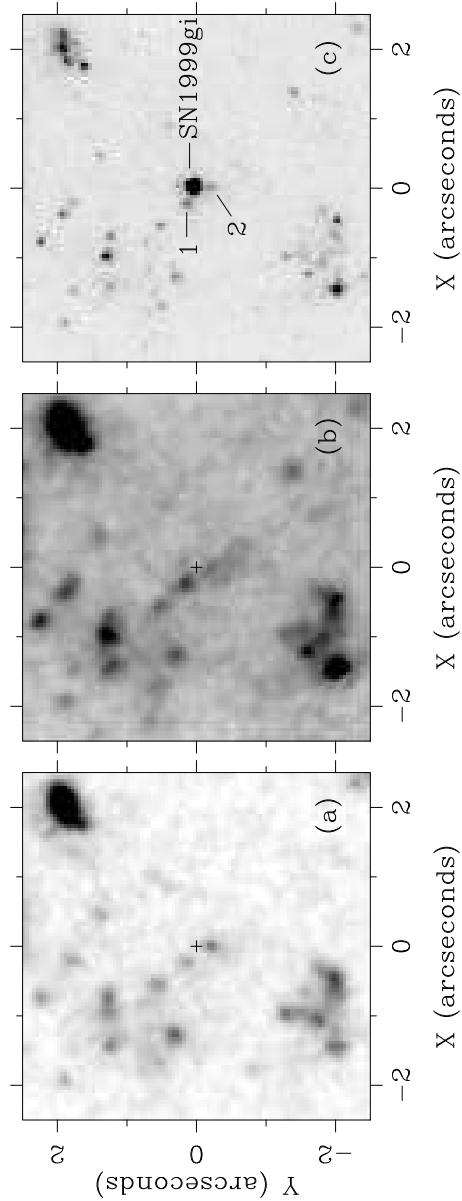}
\caption{The F300W (a) and F606W (b) images before 
the explosion of SN1999gi and the F555W (c) image of the SN in January 2001. 
The pre-explosion images are from the WF4 chip, and the post-explosion 
image was centred on the PC1 hence it has better spatial resolution. The 
position of SN1999gi is at (0,0) in all images, and marked with a cross
on the pre-explosion images. 
Many of the stars in this OB association are resolved, 
and the cluster extends approximately $20''$ across (at this
distance $0\farcs 1$ corresponds to 3.8\,pc). The two luminous
stars OB1-1 and OB1-2 are clearly visible before and after explosion, 
and both can be clearly ruled out as the progenitor. There is no 
visible object at the position of SN1999gi in either pre-explosion image.}
\end{figure}
\clearpage 

\begin{figure}
\epsscale{0.5}
\plotone{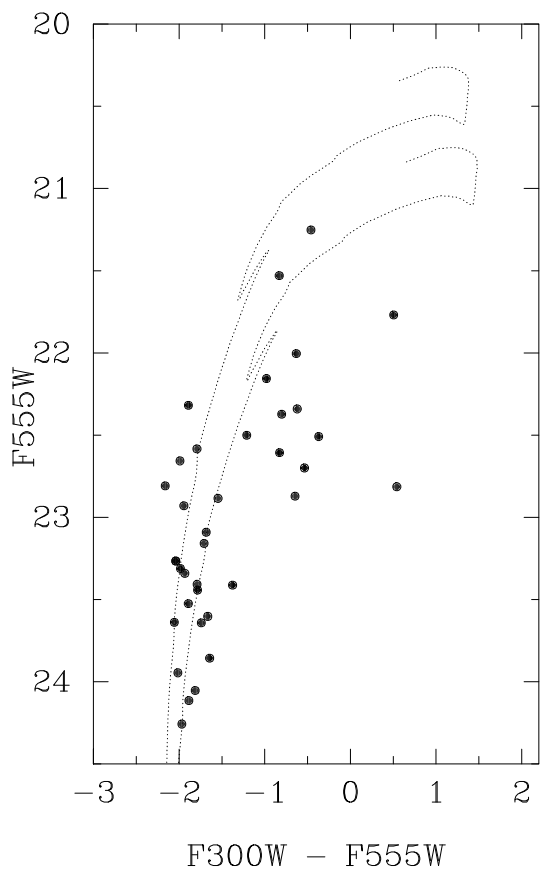}
\caption{CMD of the stars in the NGC3184-OB1 association 
which have simultaneous detections in both filters, and are within
a $25''$ diameter region (950\,pc) of SN1999gi, which 
visually appears to be the full extent of the OB1 association. The 4\,Myr 
isochrones of \cite{lej2001} are overplotted, 
reddened by E(B-V)=0.15, and 0.30 (with the 
{\sc synphot} relations). The E(B-V)=0.15 gives the best fit
given that the 0.30 track fails to match the $U_{\rm 300}$ 
brightest sources, leaving them appearing as blue stragglers.}
\end{figure}
\clearpage

\begin{figure}
\epsscale{1.0}
\plotone{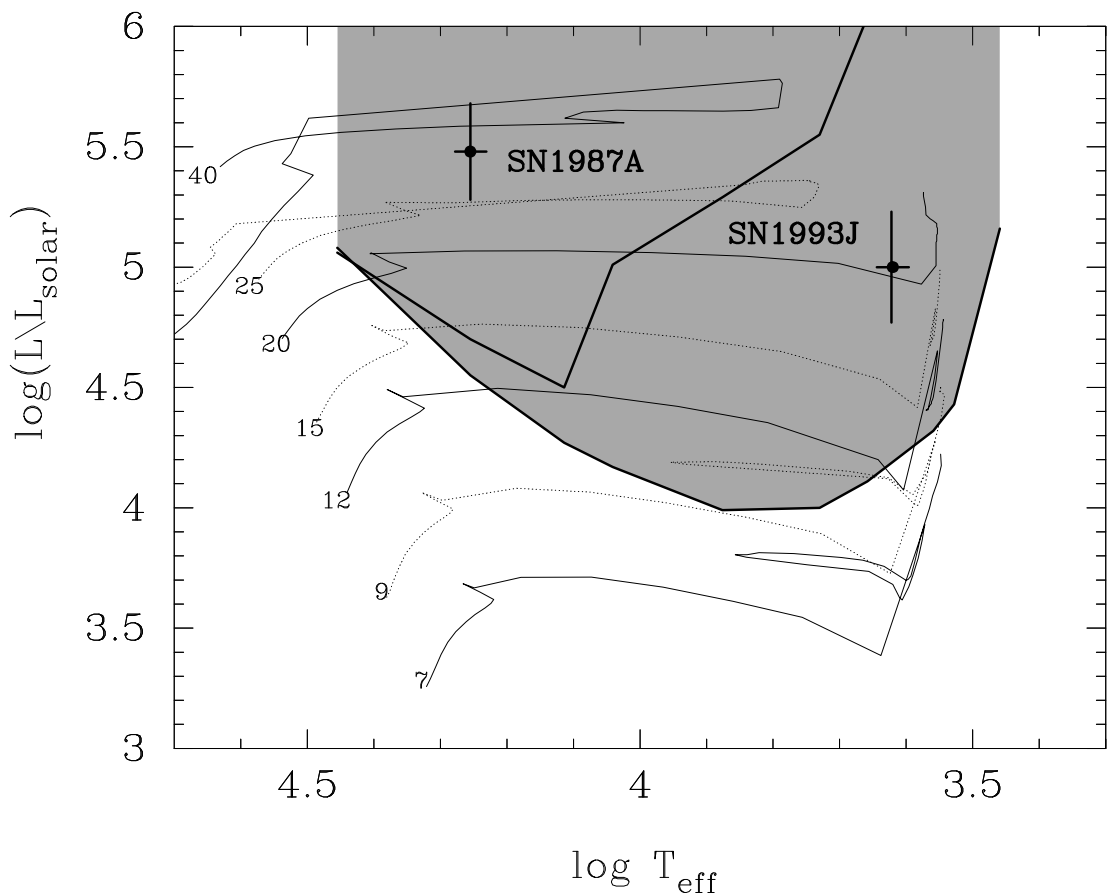}
\caption{The Geneva evolutionary tracks \citep{mey94,sch92} 
for 7$-$40\,M$_{\odot}$ plotted
with the positions of the progenitor of SN1987A and SN1993J indicated. 
The tracks are alternated between dotted and solid for clarity. 
The luminosity limits as a function of stellar effective temperature
are plotted as the thick solid lines. The WFPC2 pre-explosion
frames should be sensitive to all objects lying in the shaded region
above these lines, with the F300W filter limit being the upper curve (with
data taken from Table\,2). This indicates an upper limit to the 
progenitor of SN1999gi of 9$^{+3}_{-2}$M$_{\odot}$.}
\end{figure}
\clearpage 

%\begin{figure}
%\plottwo{f2a.eps}{f2b.eps}
%\caption{This is an example of a multipart figure with a long figure caption 
%that must be set as a paragraph.  The processor has to buffer the text of the
%caption, so it is good not to be too wordy, but that would make for
%poor communication as well.\label{fig2}}
%\end{figure}

%% If you are not including electronic art with your submission, you may
%% mark up your captions using the \figcaption command. See the 
%% No more than seven \figcaption commands are allowed per page, 
%% so if you have more than seven captions, insert a \clearpage 

\end{document}